\documentclass[prb,twocolumn,showpacs]{revtex4}

\usepackage{graphicx}
\usepackage{dcolumn}
\usepackage{bm}

\newcommand{\be}{\begin{equation}}
\newcommand{\ee}{\end{equation}}
\newcommand{\bea}{\begin{eqnarray}}
\newcommand{\eea}{\end{eqnarray}}
\newcommand{\lsim}{\raise.35ex\hbox{$<$}\kern-0.75em\lower.5ex\hbox{$\sim$}}
\newcommand{\gsim}{\raise.35ex\hbox{$>$}\kern-0.75em\lower.5ex\hbox{$\sim$}}

\begin{document}
\draft
\title{Coexistence of d$_{x^2-y^2}$-wave superconductivity and antiferromagnetism \\
induced by a staggered field}
\author{Yasuhiro Saiga and Masaki Oshikawa}
\address{Department of Physics, Tokyo Institute of Technology, Oh-okayama, Meguro-ku, 
Tokyo 152-8551, Japan}

\begin{abstract}

The two-dimensional {\it t-J} model in a staggered field is studied 
by exact diagonalization of small clusters.
For the low-hole-density region and a realistic value of $J/t$, 
it is found that the presence of a staggered field strengthens the attraction
between two holes.
With increasing field, the d$_{x^2-y^2}$-wave superconducting correlations are enhanced
while the extended-s-wave ones hardly change.
This implies that coexistence of the d$_{x^2-y^2}$-wave superconducting order
and the commensurate antiferromagnetic order occurs in a staggered field.

\end{abstract}

\pacs{74.20.-z, 74.20.Mn, 71.27.+a}


\maketitle


\section{Introduction}\label{secI}

In high-$T_{\rm c}$ cuprates,
the antiferromagnetically ordered phase and 
the superconducting phase
were separately observed
in the plane of doping concentration and 
temperature.~\cite{Dagotto,Imada-etal}
Thus antiferromagnetism and superconductivity may be
thought as competing with each other.
However,
some of the recent experiments suggest a possibility
of their coexistence, although it is still controversial.~\cite{Sidis-etal,Mook-etal-01,Hodges-etal,Stock-etal,Mook-etal-02,Breitzke-etal}

Elastic neutron scattering experiments in YBa$_2$Cu$_3$O$_y$ with $y=6.5$
and $6.6$
show
that magnetic intensity emerges near room temperature
at the momentum $(\pi,\pi)$ in units of the reciprocal lattice 
parameter.~\cite{Sidis-etal,Mook-etal-01,Mook-etal-02}
The intensity increases continuously with decreasing temperature.
Remarkably, an upturn of the intensity is observed
at a superconducting transition temperature.
This suggests coexistence of superconductivity and
the commensurate 
antiferromagnetic (AF) 
order.
Moreover, nuclear quadrupole resonance 
measurements have
revealed the presence of magnetic moments in the superconducting state on
Hg$_{0.8}$Cu$_{0.2}$Ba$_2$Ca$_2$Cu$_3$O$_{8 + \delta}$.~\cite{Breitzke-etal}
The observed magnetic moments in these materials are often regarded as
a consequence of the formation of the d-density wave
order.~\cite{Chakravarty-etal}
However, this interpretation leaves a difficulty since the d-density
wave order suppresses 
superconductivity.~\cite{Hamada-Yoshioka}
Thus we should also
consider an alternative scenario that the magnetic moments are
due to the ordered Cu spins.

So far, the possible coexistence of
antiferromagnetism and d-wave superconductivity in the two-dimensional (2D) 
{\it t-J} model
has been discussed at low hole doping, in a variational approach~\cite{Himeda-Ogata}
and quantum Monte Carlo calculations.~\cite{Sorella-etal}
The possibility of the coexistence~\cite{Inaba-etal} poses a fundamental question on
interplay between antiferromagnetism and d-wave superconductivity.
Before answering whether the coexistence actually takes place in cuprate superconductors,
we would like to clarify whether those two orders can coexist in strongly correlated
electron systems.
To clarify the matter, it would be useful
to study the hole pairing and superconductivity in a staggered
field, which forces the system to have the AF order. 
Indeed, for the 1D {\it t-J} model in a staggered field, 
the superconducting correlation was
found to be the most dominant for $J/t \sim 0.4$.~\cite{Bonca-etal,Prelovsek-etal}

In this paper,
we investigate the 2D {\it t-J} model on a square lattice in a staggered field
coupled to electron spins.
While the staggered field is introduced here as an artificial parameter
to induce the AF order, it may arise naturally
if three-dimensional
inter-plane interactions are treated in a mean field theory.
We employ exact diagonalization for the $4 \times 4$,
$\sqrt{18} \times \sqrt{18}$, $\sqrt{20} \times \sqrt{20}$, and
$\sqrt{26} \times \sqrt{26}$ clusters
with periodic boundary conditions.
Our results demonstrate that a staggered field actually enhances
the pairing of two holes and the d$_{x^2-y^2}$-wave superconductivity.

\section{Model}

We consider the following Hamiltonian given by
\bea
  {\cal H} &=& -t \sum_{\langle \vec{i},\vec{j} \rangle \sigma} \left( \tilde{c}_{\vec{i} \sigma}^{\dagger}
\tilde{c}_{\vec{j} \sigma} + {\rm H.c.} \right)
+J \sum_{\langle \vec{i},\vec{j} \rangle} \left( \vec{S}_{\vec{i}} \cdot
\vec{S}_{\vec{j}} - \frac{1}{4} n_{\vec{i}} n_{\vec{j}} \right) \nonumber \\
           & & -h \sum_{\vec{i} \in {\rm A}} S^z_{\vec{i}} + h \sum_{\vec{j} \in {\rm B}} S^z_{\vec{j}},\label{Hamiltonian}
\eea
where $\langle \vec{i},\vec{j} \rangle$ is the nearest neighbors.
The constrained fermion operator $\tilde{c}_{\vec{i} \sigma}$ is given by
$\tilde{c}_{\vec{i} \sigma} = c_{\vec{i} \sigma} (1 - n_{\vec{i},- \sigma})$,
which means that double occupancy at each site is excluded.
The last two terms are due to the presence of a staggered field whose magnitude is
denoted by $h$; A and B represent the two sublattices on a square lattice.
We refer to this model as the {\it t-J-h} model.
In this work we fix $J/t=0.4$ which is considered as a realistic value,
and 
vary $h/t$ as a parameter.

\section{Numerical Results}

We first discuss the hole correlation function given by
$C_{\rm hole} (\vec{r}) = (1/N) \sum_{\vec{i}} \:
\langle n_{\rm h} ({\vec{i}}) n_{\rm h} ({\vec{i}+\vec{r}}) \rangle$.
Here 
$N$ is the number of lattice sites,
$n_{\rm h} (\vec{i}) = 1 - n_{\vec{i}}$,
and $\langle \cdots \rangle$ denotes the expectation value in the 
zero-momentum
ground state.
In the left panel of Fig.\ 1 we show the distance dependence of $C_{\rm hole}(r)$ with
$r \equiv | \vec{r} |$ for the hole densities $N_{\rm h}/N = 2/18 \simeq 0.111$ and $4/18 \simeq 0.222$.
For two holes and $h=0$,
the most dominant correlations are at $r = \sqrt{2}$, namely,
when the holes stay at the next-nearest neighbors.~\cite{Poilblanc94,Poilblanc95}
As $h/t$ increases, correlations at the nearest neighbors ($r=1$)
become stronger than those at $r = \sqrt{2}$,
and contribution at longer distances is suppressed.
For four holes,
correlations at $r=1$ are enhanced while
ones at the largest distance hardly change.
This means that the presence of a staggered field makes the interaction between
two holes attractive but the hole-pairs are well separated.

In order to analyze the obtained data in the two-hole case, 
we calculate the root-mean-square separation
of the hole pair~\cite{Poilblanc-etal,Leung}
defined as $r_{\rm rms} \equiv \sqrt{\langle r^2 \rangle}$
where
$\langle r^2 \rangle = \sum_{\vec{r} (\ne \vec{0})} |\vec{r'}|^2 C_{\rm hole} (\vec{r})
\big/ \sum_{\vec{r} (\ne \vec{0})} C_{\rm hole} (\vec{r})$.
Here $|\vec{r'}|$ takes the 
shortest distance between two holes on the lattice
with periodic boundary conditions.
The right panel of Fig.\ 1 shows the $h/t$-dependence of $r_{\rm rms}$.
It is clear that the separation of the hole pair becomes smaller
with increasing staggered field.
For 
larger $h/t$, the value of $r_{\rm rms}$ is less affected by finite-size effects.
The present result suggests that the staggered field does help
binding of two holes, and that the hole binding survives in the
thermodynamic limit at least for a sufficiently large $h/t$.
\begin{center}
\begin{figure}
\vspace{0.3cm}
\includegraphics[width=8.4cm]{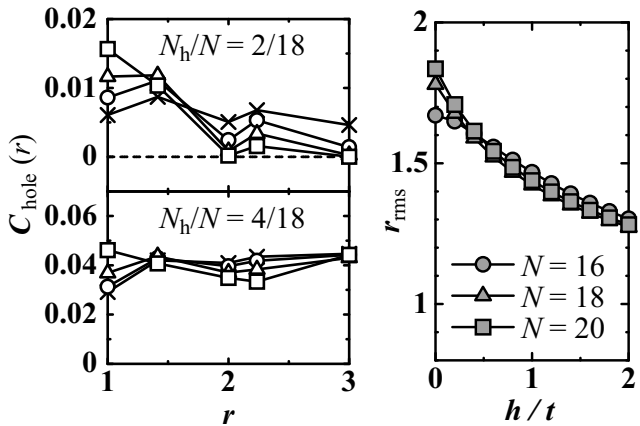}
\vspace{0.2cm}
\caption{Left: Equal-time hole correlations as a function of distance in the 2D
$t$-$J$-$h$ model with $N=18$, $J/t=0.4$, and various values of $h/t$.
Crosses, circles, triangles, and squares are the data for $h/t=0.0, 0.4, 1.0$, and $2.0$,
respectively.
Right: Root-mean-square separation of the hole pair as a function of $h/t$.
$J/t=0.4$.}
\label{holecorr}
\end{figure}
\end{center}

A tendency of hole binding can be obtained also from the binding energy,
which is given by~\cite{Riera-Young}
\be
E_{\rm B} = E_0 (N_{\rm h}=2) + E_0 (N_{\rm h}=0) - 2 E_0 (N_{\rm h}=1).
\ee
Here $E_0 (N_{\rm h})$ denotes the ground-state energy with $N_{\rm h}$ holes
in $N$ sites.
A negative value of $E_{\rm B}$ indicates the presence of hole binding.
In Fig.\ 2(a) we show the dependence of the binding energy on the staggered field $h$.
The binding energy is negative in the whole range of $h/t$ and has a peak at $h/t \sim 0.8, 1.2$, $1.2$, and $0.8$ for $N=16, 18$, $20$, and $26$,
respectively.
The obtained results apparently imply that hole 
pairing 
is suppressed
by a small staggered field.
However, being an intensive quantity, the binding energy is 
severely
affected by finite-size effects.~\cite{Poilblanc-etal}
Figure 2(b) shows the size dependence of the binding energy
for various values of $h/t$.
In fact, 
although at $h=0$ (and $J/t = 0.4$) the binding energy for $N
\le 26$ takes negative values, 
an extrapolation from the data
rather indicates 
the absence of hole binding
in the thermodynamic limit.
This has been already discussed in earlier studies~\cite{Shih-etal,Chernyshev-etal}
for $N \le 32$.
On the other hand, for larger $h/t$, we find the size dependence
to be substantially smaller.
It seems that
the binding energy
remains negative
with increasing $N$ 
in the presence of a staggered field.
Therefore there is a possibility that 
two holes in the bulk limit tend to be 
bound
even by a small field.
In particular,
for a large $h/t$ ($h/t \; \gsim \; 1.0$)
the weak size dependence of the negative binding energy strongly
suggests the hole pairing in the bulk limit.
\begin{center}
\begin{figure}
\includegraphics[width=8cm]{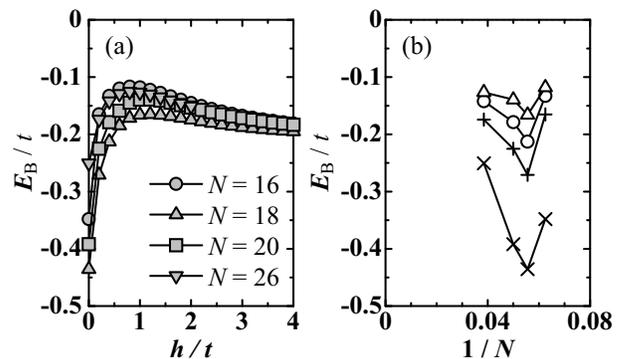}
\vspace{0.2cm}
\caption{(a) Binding energy as a function of $h/t$ in the 2D {\it t-J-h} model with $J/t=0.4$.
(b) Size dependence of the binding energy.
Crosses, pluses, circles, and triangles are the data for $h/t=0.0, 0.2, 0.4$, and $1.0$,
respectively. $J/t=0.4$.
}
\label{bindingenergy}
\end{figure}
\end{center}

The pairing of holes is also consistent with the enhanced superconducting
correlation discussed below.
We calculate the equal-time superconducting
correlations given by
$C_\alpha (\vec{r}) = (1/N) \sum_{\vec{i}} \:
\langle \Delta_\alpha^\dagger (\vec{i}) \Delta_\alpha (\vec{i}+\vec{r})
\rangle$.~\cite{Dagotto-Riera,Dagotto-etal94}
The singlet pairing operator $\Delta_\alpha (\vec{i})$ is defined as
$\Delta_\alpha (\vec{i}) = (1/\sqrt{2}) \sum_{\vec{\epsilon}} f_\alpha (\vec{\epsilon})
\left( c_{\vec{i} \uparrow} c_{\vec{i}+\vec{\epsilon},\downarrow}
- c_{\vec{i} \downarrow} c_{\vec{i}+\vec{\epsilon},\uparrow} \right)$,
where $\vec{\epsilon}$ is $(\pm 1,0)$ and $(0,\pm 1)$.
For the extended-s-wave pairing symmetry ($\alpha = {\rm s}$), 
we put $f_{\rm s} (\vec{\epsilon}) = +1$ at all $\vec{\epsilon}$.
For the d$_{x^2-y^2}$ symmetry ($\alpha = {\rm d}$), we put 
$f_{\rm d} (\vec{\epsilon}) = +1$ at $\vec{\epsilon}=(\pm 1,0)$ and
$f_{\rm d} (\vec{\epsilon}) = -1$ at $\vec{\epsilon}=(0,\pm 1)$.
Figure 3 shows the distance dependence of $C_{\rm d} (r)$
and $C_{\rm s} (r)$ for the hole densities $0.111$ and $0.222$
in $18$ sites.
For $n_{\rm h} \; \lsim \; 0.2$, with increasing field, 
the d$_{x^2-y^2}$-wave superconducting correlations are enhanced
at all distances with $r \ge 1$.
In contrast, the extended-s-wave ones hardly change,
especially at long distances.
This 
implies that the presence of a staggered field helps the d$_{x^2-y^2}$-wave
superconductivity.
\begin{center}
\begin{figure}
\includegraphics[width=8cm]{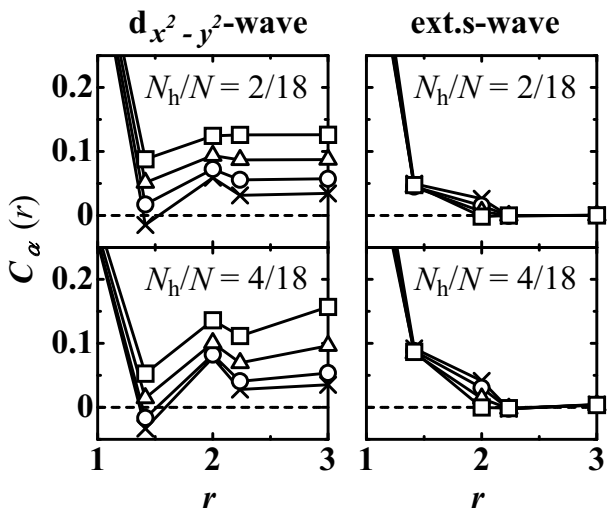}
\vspace{0.2cm}
\caption{Equal-time superconducting correlations as a function of distance
in the 2D {\it t-J-h} model with $N=18$, $J/t=0.4$, and
$h/t=0.0, 0.4, 1.0$, and $2.0$. 
The symbols are the same as in the left panel of Fig.\ 1.
}
\label{paircorr}
\end{figure}
\end{center}

Calculation of
the pair spectral function~\cite{Dagotto-etal90,Poilblanc93}
should provide
another evidence for the d$_{x^2-y^2}$-wave
pairing enhanced by
a staggered field.
The pair spectral function is given by
\bea
  P_\alpha (\omega) &=& \sum_{n} | \langle \Psi_n (N_{\rm h} = 2) | \Delta_{\alpha}^{\rm tot} | \Psi_0 (N_{\rm h} = 0) \rangle |^2 \nonumber \\
                    & & \times \delta ( \omega - E_n (N_{\rm h} = 2) + E_0 (N_{\rm h} = 0) + \mu),
\eea
where $\Delta_{\alpha}^{\rm tot} = \sum_{\vec{i}} \Delta_\alpha (\vec{i}) / \sqrt{N}$,
$\mu = E_0 (N_{\rm h} = 2) - E_0 (N_{\rm h} = 0)$,
and $| \Psi_n (N_{\rm h}) \rangle$ denotes an eigenstate with energy $E_n (N_{\rm h})$
in the $N_{\rm h}$-hole system.
The left panel of Fig.\ 4 shows the $\omega$-dependence of $P_{\rm d} (\omega)$
for various values of $h/t$.
The overall feature approximated by a Lorentzian is 
insensitive to the system size at each $h/t$.
The peak at $\omega = 0$ (i.e., the coherent peak) grows with increasing $h/t$,
which means that the pairing becomes strong.
The contribution for $\omega > 0$, which seems to be a continuum spectrum,
is relatively suppressed,
but a peak with secondary dominant intensity appears.
Concerning this peak for $h/t = 2.0$,
the values of energy $\omega$ and residue $z$ are
$(\omega / t,z)=(3.95,0.268), (3.96,0.273)$, and $(3.95,0.265)$ for $N=16, 18$, and $20$,
respectively.
The weak size dependence of both values of $\omega$ and $z$ indicates
that the secondary peak may be a delta-function contribution rather than a part of
continuum spectrum in the bulk limit.

\begin{center}
\begin{figure}
\includegraphics[width=7.5cm]{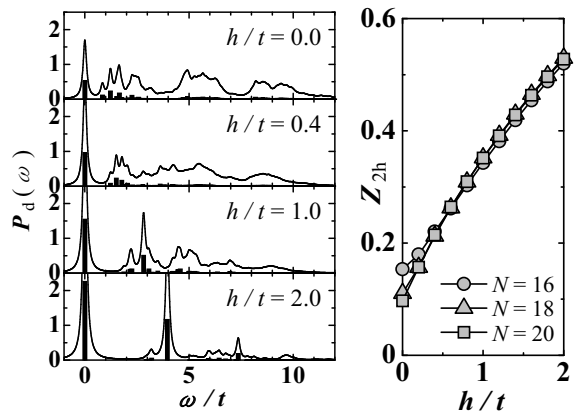}
\vspace{0.2cm}
\caption{
Left: Pair spectral function with d$_{x^2 - y^2}$-wave symmetry
in the 2D {\it t-J-h} model with 18 sites, $J/t=0.4$,
and various values of $h/t$.
The delta functions (vertical bars) are broadened by a Lorentzian with a width
of 0.1$t$ (solid curves).
Right: 
Spectral weight $Z_{\rm 2h}$ as a function of $h/t$
in the 2D {\it t-J-h} model with $J/t=0.4$.
}
\label{pairspectralfunc}
\end{figure}
\end{center}

We estimate the 
spectral weight defined as~\cite{Poilblanc93,Dagotto-Schrieffer}
\be
  Z_{\rm 2h} = \frac{| \langle \Psi_0 (N_{\rm h}=2) | \Delta_{\rm d}^{\rm tot} |
  \Psi_0 (N_{\rm h}=0) \rangle |^2}
  {\langle \Psi_0 (N_{\rm h}=0) | (\Delta_{\rm d}^{\rm tot})^\dagger \Delta_{\rm d}^{\rm tot} |
  \Psi_0 (N_{\rm h}=0) \rangle},\label{Z2h}
\ee
which corresponds to the coherent peak of $P_{\rm d} (\omega)$ at
$\omega=0$.
Note that $Z_{\rm 2h}$ is between 0 and 1
because the denominator of Eq.\ (\ref{Z2h}) is equal to integration of
$P_{\rm d}(\omega)$ over $\omega$.
The right panel of Fig.\ 4 shows $Z_{\rm 2h}$ as a function of $h/t$.
The weight is monotonically increasing function of $h/t$.
Again, the size dependence of $Z_{\rm 2h}$ is weak for a large $h/t$.
Therefore we expect that the
coherent peak survives in the
thermodynamic limit for a sufficiently large $h/t$.

While our results suggest the enhancement of hole pairing by a
staggered field,
it is possible that the attraction between holes leads to phase
separation.
We calculate the clustering energy given by~\cite{Riera-Young}
\be
E_{\rm C} = E_0 (N_{\rm h}=4) + E_0 (N_{\rm h}=0) - 2 E_0 (N_{\rm h}=2).
\ee
If this quantity is negative, the phase separation is expected to occur.
The results for $N=18$ and $20$ are shown in Fig.\ 5,
which suggests that
the region $0 \le h/t \; \lsim \; 2$ is not interrupted
by the phase separation.~\cite{comment}
\begin{center}
\begin{figure}
\includegraphics[width=4cm]{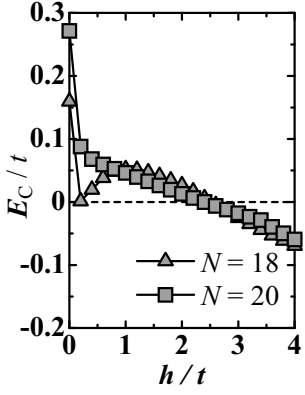}
\vspace{0.2cm}
\caption{Clustering energy as a function of $h/t$
in the 2D {\it t-J-h} model with $J/t=0.4$.}
\label{clusteringenergy}
\end{figure}
\end{center}


The staggered field should induce a finite magnetic moment on each site,
and consequently the commensurate AF long-range order.
In Fig.\ 6 we show the result on the staggered-spin correlations given by
$C_{\rm spin} (\vec{r}) = (1/N) \sum_{\vec{i}} \: (-1)^{r_x + r_y}
\langle S^z_{\vec{i}} S^z_{\vec{i}+\vec{r}} \rangle$
with $\vec{r} = (r_x,r_y)$.
Indeed, the correlations seem to remain finite in the long-distance
limit for $h/t>0$, as expected.
An important point
is that {\it both} the d$_{x^2-y^2}$-wave superconducting
correlations and the staggered-spin ones are enhanced by a staggered field.
Therefore we expect the simultaneous
presence of the d$_{x^2-y^2}$-wave superconducting
order
and the commensurate AF order in a certain range of $h/t$.
\begin{center}
\begin{figure}
\vspace{0.2cm}
\includegraphics[width=8cm]{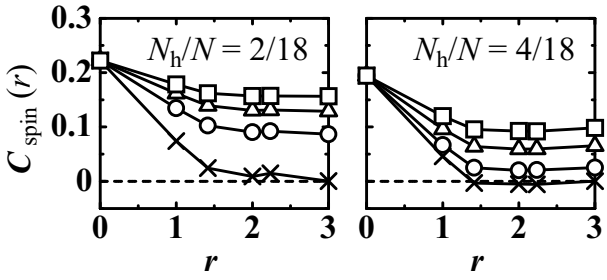}
\vspace{0.2cm}
\caption{Equal-time staggered-spin correlations as a function of distance 
in the 2D $t$-$J$-$h$ model with $N=18$, $J/t=0.4$, and 
$h/t=0.0, 0.4, 1.0$, and $2.0$.
The symbols are the same as in the left panel of Fig.\ 1.}
\label{spincorr}
\end{figure}
\end{center}

\section{Perturbation from Large-$\bm{h}$/$\bm{t}$ Limit}

Why does the presence of a staggered field help the pairing formation?
This may be understood from the large-$h/t$ limit.
In order to
treat analytically, we consider
the $t$-$J_z$-$h$ model
where the isotropic Heisenberg term ($J$) in Eq.\ (\ref{Hamiltonian})
is replaced by the Ising term ($J_z$),
following Refs.\ 14 and 15.
For $h/t \gg 1$ we can regard the single-particle hopping term ($t$)
as a perturbative one.
In the Hilbert space with all spins along the direction of the staggered field,
the second-order perturbation leads to the effective Hamiltonian given by
${\cal H}_{\rm eff} = P \tilde{\cal H}_{\rm eff} P$ where
\bea
  && \tilde{\cal H}_{\rm eff} = J_z \sum_{\langle \vec{i},\vec{j} \rangle} 
  \left( S_{\vec{i}}^z S_{\vec{j}}^z - \frac{1}{4} n_{\vec{i}} n_{\vec{j}} \right) \nonumber \\
  && - \tilde{t} \sum_{\langle \vec{i},\vec{j},\vec{\ell} \rangle \sigma} \left[ \tilde{c}_{\vec{i} \sigma}^\dagger ( 1 - n_{\vec{j}} ) \tilde{c}_{\vec{\ell} \sigma} + ( 1 - n_{\vec{i}} ) ( 1 - n_{\vec{j}} ) n_{\vec{\ell} \sigma} + {\rm H.c.} \right] \nonumber \\
  && + ({\rm other} \; \; {\rm terms}).
  \label{effectiveHamiltonian}
\eea
Here $\langle \vec{i},\vec{j} \rangle$ and $\langle \vec{i},\vec{j},\vec{\ell} \rangle$ are the nearest neighbors,
and $P = \prod_{\vec{i} \in {\rm A}} (1 - n_{\vec{i} \downarrow})
\prod_{\vec{j} \in {\rm B}} (1 - n_{\vec{j} \uparrow})$.
The second term in the right-hand side of
Eq.\ (\ref{effectiveHamiltonian}) includes hopping of a hole
which occurs only if there is another hole in the neighboring site.
Namely, it gives hopping of a hole pair.
This is generated in the second-order processes shown in Fig.\ 7.
In the 2D case the pair-hopping integral $\tilde{t}$ is given by
$t^2 / ( h + 3J_z / 2 )$ where the denominator indicates the energy difference
between the initial 
state and the intermediate one
[see Fig.\ 7(IIa) and 7(IIb)].
We note that $\tilde{t}$ is replaced by $t^2 / ( h + J_z / 2 )$ in the 1D case
[see Fig.\ 7(I)].~\cite{Bonca-etal,Prelovsek-etal}
\begin{center}
\begin{figure}
\vspace{0.2cm}
\includegraphics[width=7.5cm]{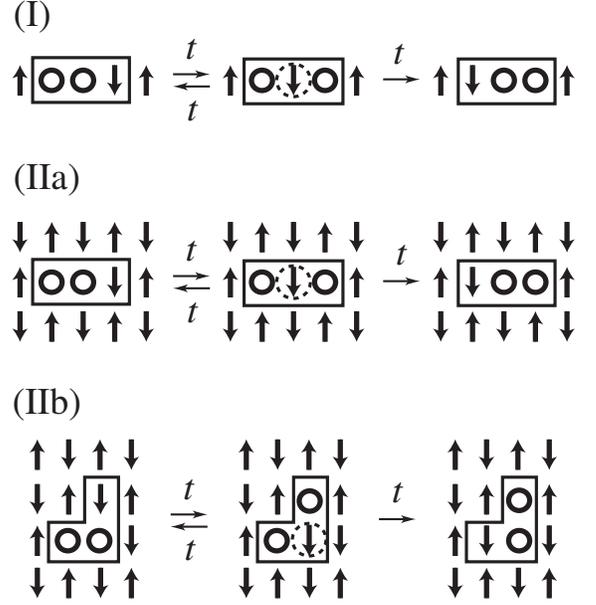}
\vspace{0.2cm}
\caption{Typical processes of the second-order perturbation
in the $t$-$J_z$-$h$ model for $h/t \gg 1$.
(I) is the 1D case, while (IIa) and (IIb) are the 2D case.
The spin surrounded by a broken circle is against the direction of the
staggered field.
The three sites surrounded by a box correspond to
$\langle \vec{i},\vec{j},\vec{\ell} \rangle$
in the second term of Eq.\ (\ref{effectiveHamiltonian}).}
\label{process}
\end{figure}
\end{center}

The fact that the effective Hamiltonian includes the hole-pair hopping
implies that hole binding and superconductivity may be naturally
realized in the antiferromagnetic background forced by the staggered
field.
The effective pair-hopping integral $\tilde{t}$ decreases as $h/t$
increases.
This suggests that a too large staggered field makes the hole pairs less
mobile,
which may be related to the phase separation at large $h/t$ discussed
above.
On the other hand, it does not mean that a smaller $h/t$ is better for
superconductivity,
because the calculation is based on the perturbation theory in $t/h$.

\section{Summary and Discussion}

We have investigated the binding energy and
various correlation functions for the 2D {\it t-J} model in a staggered field.
For the low-hole-density region at $J/t=0.4$,
the presence of a staggered field strengthens
the attraction between two holes and helps the d$_{x^2-y^2}$-wave superconductivity.
This implies that the commensurate antiferromagnetic order and
d$_{x^2-y^2}$-wave superconductivity can coexist in a strongly
correlated electron system in two dimensions.

One may ask whether calculation of small clusters with $N \sim 20$ provides some
conclusive statements in a model.
In fact, for the 2D {\it t-J} model without a staggered field,
binding effects for $N \sim 20$ can be different from those for much larger 
size.~\cite{Chernyshev-etal,BM,WS}
However, the presence of a staggered field makes the coherence length
(i.e., the size of a Cooper pair) small,
and therefore the data for $N \sim 20$ is likely 
to reach the bulk limit for a sufficiently large staggered field, 
as evidenced by the weak size dependence.
Thus our conclusion 
regarding the coexistence of superconductivity and antiferromagnetism 
should hold at least near the boundary
of the phase separation ($h/t \sim 2$).~\cite{phaseseparation}
This would be of a conceptual interest, although such a large staggered field
seems unrealistic.
An open question is whether the picture for such a large field 
connect continuously to that for smaller field.
For a realistic application of the present model,
perhaps we need to know the effect of a small staggered field,
as the effective staggered field produced by the interlayer coupling would be tiny.
Unfortunately, for a small staggered field, the size dependence is still large
and we cannot draw a definite conclusion from our present study based on small clusters.
We hope future studies to clarify this question and its relation to the experiments.

\acknowledgments

We thank G. Misguich for calling our attention to recent experiments on
high-$T_{\rm c}$ cuprates.
Y.S. is supported by JSPS Research Fellowships for Young Scientists.
The present work is supported in part by Grant-in-Aid for Scientific
Research from MEXT of Japan.



\begin{thebibliography}{99}
\bibitem{comment}
The clustering energy for $N=18$ has a dip at $h/t \sim 0.2$,
but this is likely to be due to finite-size effects.
\bibitem{phaseseparation}
It has been discussed in some models of correlated electrons
that superconductivity tends to emerge near phase separation
[see Ref.\ 24].
\end{thebibliography}
\end{document}